\documentclass[twocolumn]{aastex631}

\hypersetup{
    colorlinks=true,
    linkcolor=blue,
    filecolor=cyan,      
    urlcolor=magenta,
    citecolor=blue,
}
\usepackage{natbib}
\usepackage{hyperref}
\usepackage{amsmath}
\usepackage{array}
\usepackage{graphicx}
\usepackage{xcolor}
\usepackage{txfonts}
\usepackage{float}
\usepackage{microtype}
\usepackage{subfigure}
\usepackage{multirow}
\usepackage{rotating}
\usepackage{balance}
\usepackage[mathlines]{lineno}

\chardef\us=`\_
\newcommand{\Rs}{\(\mathrm{R}_\odot\)}

\submitjournal{ApJ}

\begin{document}


    \title{Linking Critical Heights in Solar Active Regions with 3D CME Speeds:\\ Insights from Automated and Manual PIL Detection Methods}

\author[0000-0003-4772-2492]{Harshita Gandhi}
\affiliation{Department of Physics, Aberystwyth University, Aberystwyth, Wales, UK}

\author[0000-0001-7927-9291]{Alexander W James}
\affiliation{Mullard Space Science Laboratory, University College London, Holmbury St. Mary, Dorking, Surrey, RH5 6NT, UK}

\author[0000-0002-6547-5838]{Huw Morgan}
\affiliation{Department of Physics, Aberystwyth University, Aberystwyth, Wales, UK}

\author[0000-0002-0053-4876]{Lucie Green}
\affiliation{Mullard Space Science Laboratory, University College London, Holmbury St. Mary, Dorking, Surrey, RH5 6NT, UK}



\begin{abstract}
In a space weather context, the most geoeffective coronal mass ejections (CMEs) are fast CMEs from Earth-facing solar active regions. These CMEs are difficult to characterize in coronagraph data due to their high speed (fewer observations), faintness, Earthward orientation (halo CMEs), and disruptions from associated high-energy particle storms. Any diagnostic aiding in early CME speed identification is valuable. This study investigates whether the 3D speeds of 37 CMEs are correlated with the critical heights of their source regions, to test the hypothesis that if the critical height is located at a higher altitude in the corona, the weaker magnetic field environment will enable a faster CME to be produced. Critical heights near CME onset are calculated by identifying polarity inversion lines (PIL) in magnetogram data using automated and manual methods. 3D speeds are determined by fitting a Graduated Cylindrical Shell (GCS) model to multi-viewpoint coronagraph images.  For the automated
method, we find a high correlation of 71\% $\pm$ 8\% between CME speed and critical height, dropping to 48\% $\pm$ 12\% when using CME plane-of-sky speeds, on which most previous similar studies are based. An attempt to improve the critical height diagnostic through manual PIL selection yields a lower correlation of 58\% $\pm$ 13\%. The higher correlation from the automated method suggests that encompassing the full PIL structure is a better measure of the magnetic conditions that influence CME dynamics. Our results highlight the potential for critical height as a continuously computable diagnostic for forecasting the 3D speeds of Earth-directed CMEs. 
\end{abstract}



\section{Introduction}
\label{sec:intro}

Solar phenomena, particularly coronal mass ejections \citep[CMEs;][]{hundhausen1999coronal,forbes2000review}, play a crucial role in shaping space weather, impacting both terrestrial and space-based technologies \citep{schwenn2006space,pulkkinen2007space}. Although CMEs typically take 1-3 days to reach Earth, current forecasting capabilities are limited to post-eruption periods. This limitation is especially pronounced for CMEs, whose Earth-directed trajectory makes their early detection and speed estimation challenging in near-real-time coronagraph data \citep{gopalswamy2007geoeffectiveness}. Given that fast-moving CMEs can reach Earth in less than a day, understanding pre-eruptive phases and identifying predictive parameters in active regions is crucial for improving lead times and advancing space weather forecasting \citep{gopalswamy2007geoeffectiveness,alexakis2019statistical,taylor2020space,zhang2021earth, besliu2021geoeffectiveness}.

CMEs result from the sudden destabilization of a coronal magnetic field region, following an extended phase of magnetic stress accumulation and free magnetic energy build-up within active regions \citep{forbes2006cme, schmieder2015flare, chen2017physics, majumdar2022variation}. This energy build-up culminates in the eruption phase, where, under conditions of insufficient downward magnetic tension from overlying fields, plasma and magnetic fields are expelled into interplanetary space. Observations from solar telescopes equipped with remote-sensing instruments, such as coronagraphs, Extreme ultraviolet (EUV) and X-ray imagers, spectrographs, and magnetographs, provide data across these phases, while theoretical models based on magnetohydrodynamics (MHD) offer insight into the complex physics governing CME initiation and evolution. 

The current understanding of CME initiation mechanisms divides them into resistive (e.g., magnetic reconnection) and ideal MHD instabilities, such as the torus instability \citep{kliem2006torus}. Torus instability occurs when a toroidal-shaped magnetic structure, such as a flux rope, experiences an outward hoop force that overcomes the inward magnetic tension from the overlying field \citep{deng2017roles}. Following this understanding of CME initiation mechanisms, the concept of the magnetic decay index \citep{kliem2006torus}, $n$, emerges as a key theoretical tool in the study of CME dynamics. It is a measure of quantifying how the strength of the poloidal component of an external magnetic field, B$_{ext,p}$ (the component that contributes to the tension force), changes with the radial distance, $R$, measured from the photosphere. It is defined as: 

\begin{equation}
    n = -\frac{d \ln B_{\text{ext,p}}}{d \ln R}
    \label{eq:decay_index}
\end{equation}

A structure becomes susceptible to rapid expansion and may erupt as a CME if it exists within an external magnetic field where the decay index exceeds a critical threshold, $n_{c}$ = 1.5, a principle initially established by \cite{bateman1978mhd} and further explored by \cite{kliem2006torus} in the context of CME dynamics. This critical value $n_{c}$ for torus instability can vary depending on the specific characteristics of CME source regions. This variation affects CME dynamics, as regions with a higher decay index are typically associated with faster CMEs due to the stronger outward magnetic force overcoming the confining magnetic tension. \citep{liu2007speed,xu2012relationship,deng2017roles,cheng2020initiation}. Subsequent theoretical investigations have identified ranges of critical decay index of 1.1 to 1.3 for linear flux ropes and 1.5 to 1.9 for circular configurations, with empirical studies confirming average critical values around 1.2±0.2 and 1.6±0.1 for quiescent filaments and active channel flux ropes \citep{fan2007onset,demoulin2010criteria,fan2010eruption,cheng2020initiation}
 
The critical height is the altitude above the photosphere where the decay index reaches its critical value $n_{c}$, typically around 1.5 for torus instability. At this height, the restraining forces from the overlying magnetic fields become insufficient to confine the outward magnetic pressure gradient, enabling a CME eruption. The initial acceleration of a CME is strongly influenced by the magnetic conditions at this height \citep{patsourakos2010toward,james2022evolution}, with the CME’s speed closely linked to the magnetic environment of its eruption site \citep{macqueen1983kinematics,zhang2006statistical}. We hypothesize that higher critical heights are correlated with faster CME speeds because they correspond to source regions with weaker coronal magnetic fields above the torus instability onset that allow flux ropes to accelerate more effectively after reaching the threshold ($n_{c}(h)$=1.5). When the onset of the torus instability occurs higher up in the corona (i.e. from a higher critical height), the erupting flux rope must travel a shorter distance through the weak upper corona on its way into the heliosphere and therefore experiences less retardation from magnetic tension than a CME that erupts from lower down in the corona. This will allow the flux rope to accelerate more effectively after reaching the torus instability threshold.. 

Calculating critical height from magnetic field extrapolations significantly bridges the observational gap in early CME development, where visible signs of instability are not yet apparent \citep{zuccarello2015critical}. The importance of determining the critical height lies in its ability to provide a quantifiable metric for assessing the likelihood of CME eruptions \citep{james2022evolution}. In this context, critical height serves as a valuable marker derived from extrapolated magnetic field configurations above the photosphere, identifying regions where the axis of the flux rope is at the critical height.

The complexity of CME source regions, alongside the critical height and decay index, plays a significant role in determining CME speed and dynamics. Studies show that CME speed — a crucial factor for geoeffectiveness, correlates with the magnetic properties of the source region. Active regions classified under the Hale system \citep{hale1919magnetic} as $\beta$$\gamma$$\delta$ sunspot groups, indicating complex magnetic configurations, high magnetic shear and multiple polarity inversion lines have been linked to more energetic solar events, including faster CMEs \citep{su2007determines,sammis2000dependence,wang2007comparative,falconer2002correlation,schrijver2007characteristic,georgoulis2007quantitative} \citep{srivastava2004solar,gopalswamy2006coronal,kontogiannis2019photospheric}.

Faster Earth-directed CMEs lead to stronger geomagnetic disturbances due to their higher ram pressure and shock-driving potential \citep{willis1964sudden, srivastava2004solar, siscoe2006cme, gopalswamy2008solar, zhang2021earth}. However, the speed estimation for such CMEs is subject to large uncertainties. This is primarily because faster CMEs have reduced observational data from coronagraphs, which limits the accuracy of tracking their propagation paths and velocities through interplanetary space.

Despite these advances, many past studies have relied on plane-of-sky speeds derived from single-viewpoint observations, which are subject to projection effects and often underestimate the 3D (true) speed of CMEs (e.g., \citet{burkepile2004role, richardson2010near, bidhu2017cme, pant2021investigating} and references therein). Projected speeds do not capture the 3D perspective and can result in significant errors when used to correlate CME dynamics with source region characteristics \citep{webb2000understanding, temmer2009cme,gandhi2024correcting}. 3D speed refers to the three-dimensional propagation speed of the CME, accounting for its full spatial motion rather than the apparent motion confined to the plane of the sky. Note that in this section and elsewhere, the GCS fitted CME parameters are referred to as '3D', although this is, of course, subject to the assumption that the chosen geometrical model is the correct choice of geometry for all CMEs analysed. 

The Graduated Cylindrical Shell (GCS) model developed by \citet{thernisien2006modeling,thernisien2009forward} is a widely used geometrical framework that reconstructs CME geometry in three-dimensions by combining multi-viewpoint observations from Solar TErrestrial RElations Observatory (STEREO) \citep{kaiser2008stereo} and the Large Angle and Spectrometric Coronagraph (LASCO) \citep{brueckner1995large} instruments. This approach mitigates projection effects and provides a more accurate representation of CME dynamics. For a detailed review of projected speeds, refer to \citet{gandhi2024correcting}, where they compared plane-of-sky and 3D speeds of 360 CMEs, demonstrating that the average 3D speed is approximately 1.3 times greater than the plane-of-sky speed. These findings underscore the importance of using 3D speed measurements to achieve reliable correlations with predictive parameters like critical height. By obtaining 3D speeds, we can better understand how source region conditions relate to 3D CME speeds.

In this study, we test this hypothesis by examining the relationship between the 3D speeds of 37 CMEs, derived from the GCS geometrical model, that originate from active regions with varying Hale classifications, and the critical heights calculated at CME onset.  A unique aspect of this study is the use of both automated and manual methods for detecting PILs and calculating critical heights, providing a comprehensive comparison of the two approaches. The automated method enables efficient and continuous tracking of critical height, suitable for real-time forecasting, while the manual method allows precise post-event analysis in complex regions. By examining the critical height along PILs and its influence on CME dynamics, our study clarifies how these parameters can serve as reliable indicators of CME speed.

Section \ref{sec:method} details the data selection, methodology for automated and manual PIL detection, critical height calculations, and the 3D reconstruction of CME speeds. The uncertainties associated with these measurements are carefully quantified, adding robustness to the results. Section \ref{sec:Results} presents the results from automated and manual PIL methods. Section \ref{sec:D&C} then provides a comprehensive discussion, comparing the strengths and limitations of each method, followed by the conclusions.

\section{Data and Method}
    \label{sec:method}
    
This section describes the selection and analysis of 37 active regions of different hale classifications that produced one or more CMEs during the time period studied. The primary objective of this study is to investigate the relationship between differences in the extrapolated coronal magnetic field above an active region and the kinematics of CMEs. 

The following criteria were used to select the active regions: Active regions were required to be within $\pm$45 degrees longitude of disk center. We reviewed observations using JHelioviewer\footnote{https://www.jhelioviewer.org/}  \citep{muller2009jhelioviewer,muller2017jhelioviewer} to observe CME signatures below the occulting disk, such as flare signatures or loop structures within the active region. These observations were correlated with the appearance of CMEs in the LASCO C2 coronagraph to estimate the approximate onset time of the CMEs. 

Section \ref{sec:PILDCH} details the method for detecting polarity inversion lines and determining the critical height. Section \ref{sec:VE} describes the data and the methodology used for fitting CMEs using the GCS technique from multiple viewpoints, followed by the estimation of height-time profiles and 3D speeds.

\subsection{PIL Detection and Critical Height Estimation}
    \label{sec:PILDCH}

\begin{figure}[ht]
     \centering
     \begin{subfigure}
         \centering
         \includegraphics[width=0.46\textwidth]{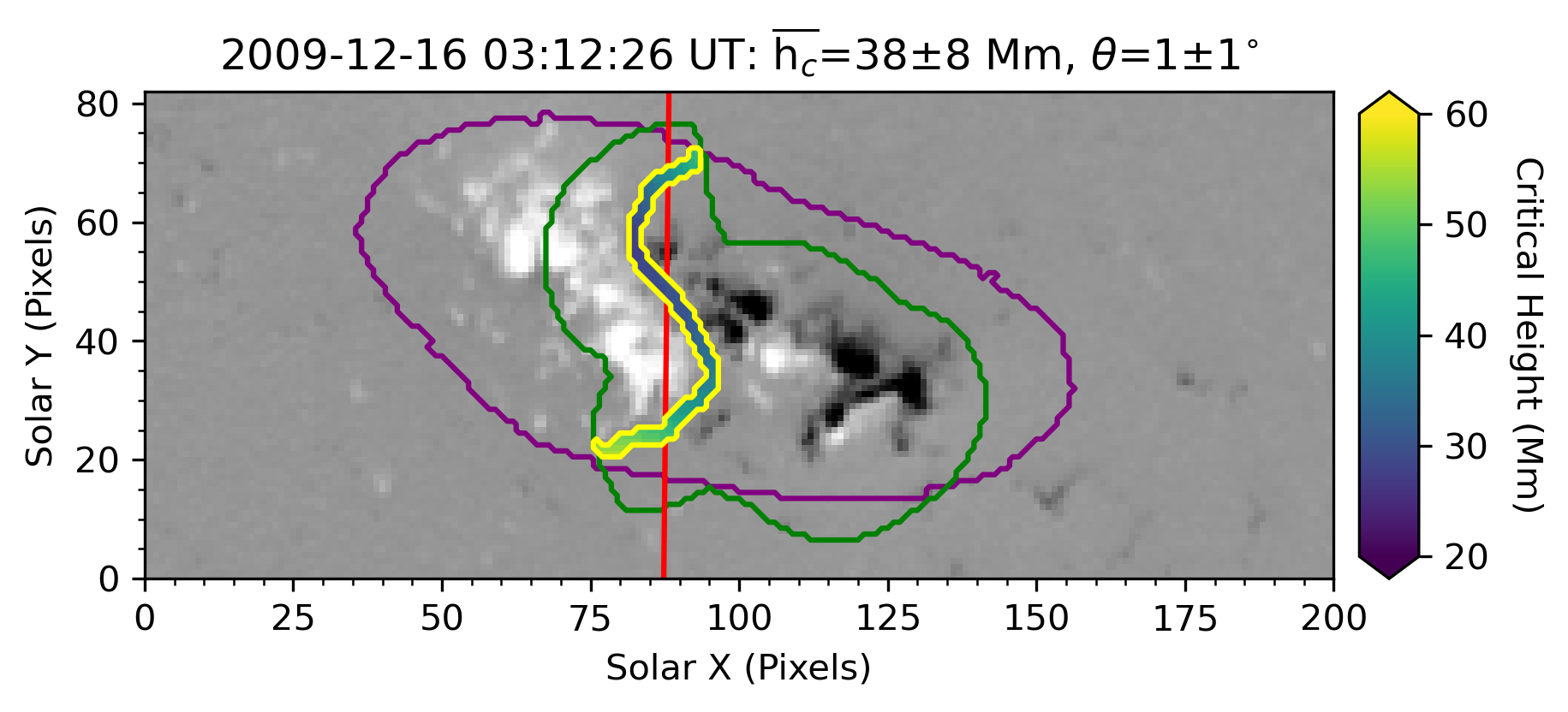}
     \end{subfigure}
     \hfill
     \begin{subfigure}
         \centering
         \includegraphics[width=0.46\textwidth]{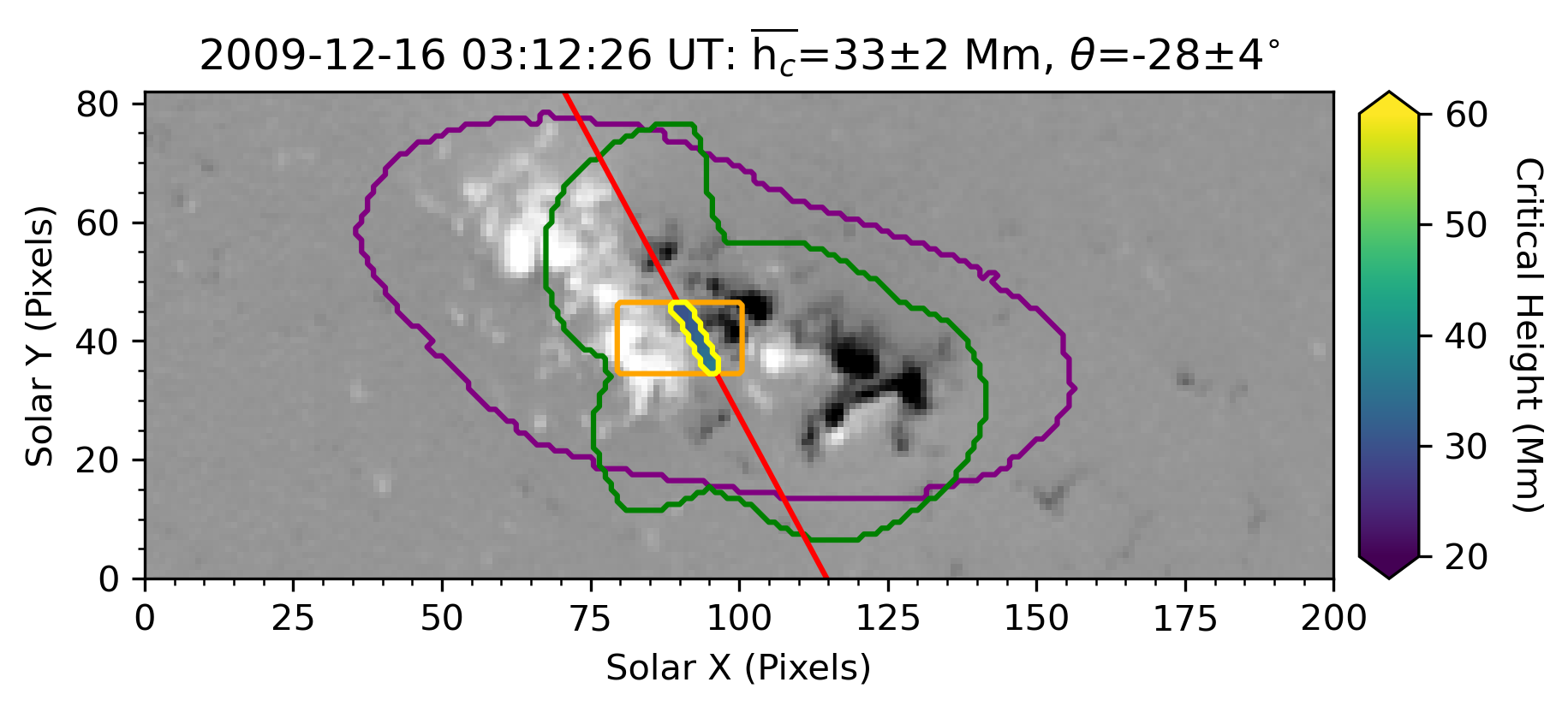}
     \end{subfigure}
        \caption{\textcolor{black}{NOAA AR 11035 within TARP 13189. Positive (negative) magnetic flux is shown in white (black). One pixel represents about 1.4 Mm. The purple contour outlines the weak field bitmap region, and the green contour outlines a region where strong positive and negative polarities are close to each other. In the lower panel, an additional orange box defines a sub-region where EUV observations suggest a CME erupted from. The longest continuous PIL section within all relevant contours is identified (contoured in yellow), and each pixel along it is coloured to represent the critical height above it. A linear fit to each PIL is shown in red and used to infer the tilt angle of each PIL relative to the Y axis (solar north).}}
    \label{fig:pil}
\end{figure}

To model the coronal magnetic field of each active region, we use the Cartesian, Fourier method of \citet{alissandrakis1981field} to extrapolate the potential magnetic field above radial field magnetograms from the Space-Weather HMI Active Region Patch (SHARP; \citealp{bobra2014sharps}) data series. 
We also use some Space-Weather MDI Active Region Patches (SMARPs; \citealp{bobra2021smarps}), but this dataset does not contain radial field magnetograms. 
Instead, we approximate the radial field component using the method of \citet{Leka2017radial}, first extrapolating the full vector potential field from their line-of-sight field components and then taking the radial field component from the bottom layer of the extrapolation volume. 
Critical heights based on potential field extrapolations of MDI magnetograms that are radialised in this way generally agree well with critical heights obtained from extrapolations of HMI radial-field observations of the same active regions \citep{james2024cycle}. The potential field is commonly used as an approximation for the field external to a current-carrying magnetic flux rope \citep[e.g.][]{torok2007simulations,zuccarello2015critical,wang2017critical}, and the extrapolated horizontal field is used as a proxy for the poloidal field component \citep[e.g.][]{liu2008instabilites,zuccarello2015critical,wang2017critical,james2022evolution}, which is relevant to the decay index. Under these assumptions, we calculate the decay index throughout each extrapolated coronal field volume as given by Equation \eqref{eq:decay_index}. We use two slightly different methods to estimate the critical height above polarity inversion lines (PILs) in active regions at the onset time of each CME: an automated method (Method I) and a manual method incorporating additional observations (Method II), both of which are based on the method used by \citet{james2024cycle}.

In both methods, we locate photospheric PILs by spatially smoothing magnetograms using a $20 \times 20$ pixel moving average window and identifying the boundaries between positive and negative magnetic fluxes. Then, we use ``weak field’’ bitmaps of magnetic field strength from the SHARP/SMARP data series to select only PIL pixels that are located in areas where the magnetic field is stronger than the quiet Sun within the purple contours in Figure \ref{fig:pil}. Furthermore, we select only PIL pixels situated between strong opposite magnetic polarities using a calculation similar to that of the \citet{schrijver2007R} $R$ parameter within the green contours in Figure \ref{fig:pil}. These techniques can result in the detection of multiple disconnected PIL sections. 
\citet{james2024cycle} use all detected PIL pixels to calculate the critical height, but here, we select only the longest continuous section for further analysis.

For the other method (Method II), we take the extra step of also using EUV observations of flare arcades, flare ribbons, and coronal dimmings to identify the specific part of each active region that was active during the eruption of each CME. Within this sub-region (orange box in the lower panel of Fig. \ref{fig:pil}), we use the previously described steps to search for the PIL section that was most relevant to the eruption. We note that sometimes this PIL is in a different part of the active region to the longest PIL identified by the first method. An example from each PIL detection method is shown in Fig. \ref{fig:pil}.

Once each PIL is identified, we examine the column in the extrapolation volume above each PIL pixel and note the height at which the critical decay index ($n_{\mathrm{c}} = 1.5$) occurs. If more than one critical height is found above a given PIL pixel \citep[e.g. there is a ``saddle’’ in the decay index-height profile,][see the bottom right panel of Fig. \ref{fig:decay_index} for an example]{guo2010confined,luo2022saddles}, we select the lowest critical height because they tend to be more comparable to the critical heights in regions that do not exhibit saddles, and CMEs have been observed to initiate from these lower critical heights despite sitting beneath a second torus-stable zone \citep{wang2017critical}. Finally, we take the mean of the critical heights found above all pixels along the PIL to obtain a single critical height representative of each magnetogram. We give the standard deviation of critical heights along the PIL as an estimate of error.

\begin{figure}[ht]
    \centering
        \includegraphics[width=0.45\textwidth]{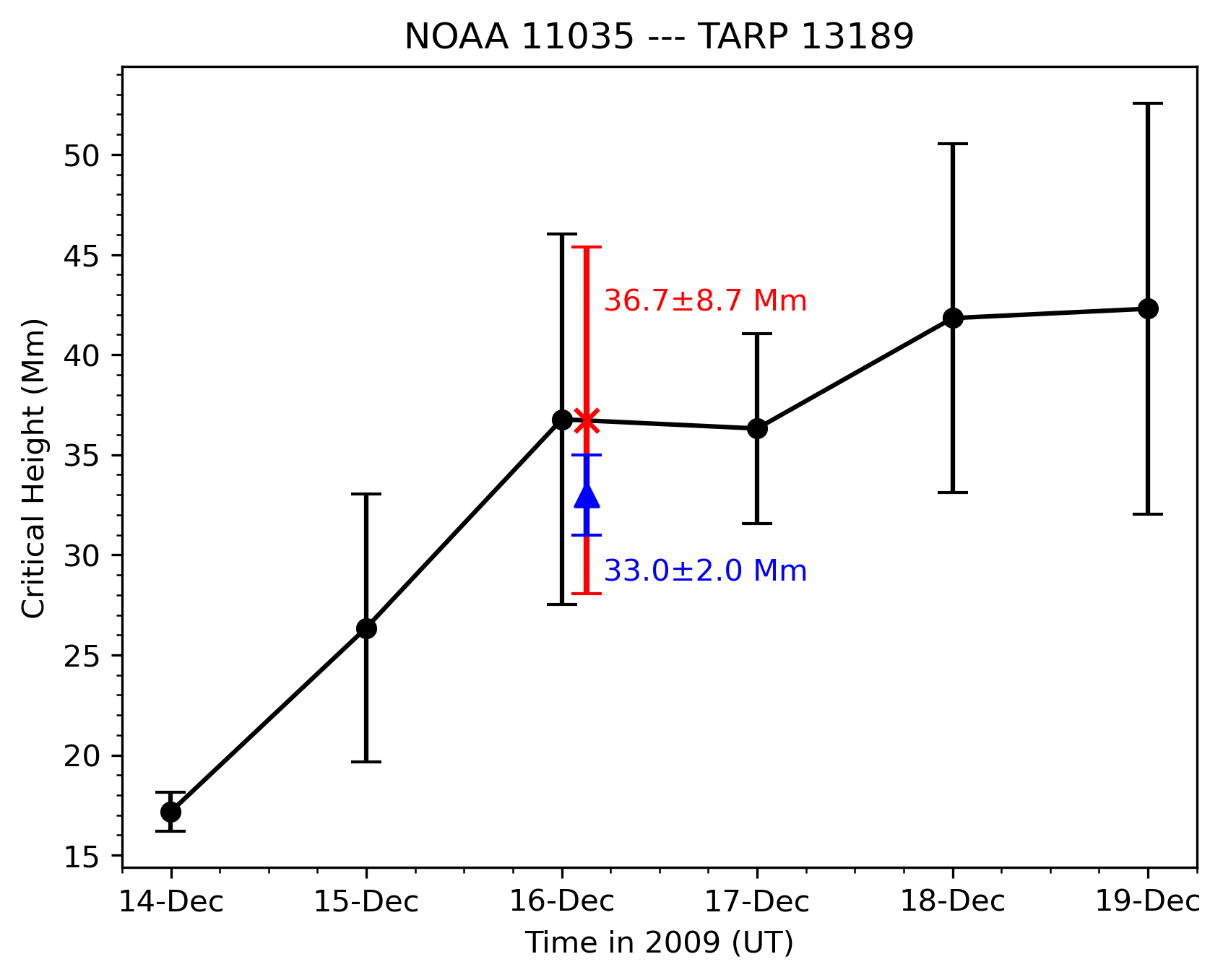}
    \caption{Temporal variation of the critical height (Mm) for NOAA AR 11035 from December 14 to 19, 2009. Black error bars show measurement uncertainties (standard deviation along the PIL). The red `X' marks the interpolated critical height at CME onset, while the blue triangle indicates the critical height from the second method for the CME-associated PIL subsection.}
    \label{fig:CH}
\end{figure}

\begin{figure*}[htp]
     \centering
     \begin{subfigure}
         \centering
         \includegraphics[width=0.73\textwidth]{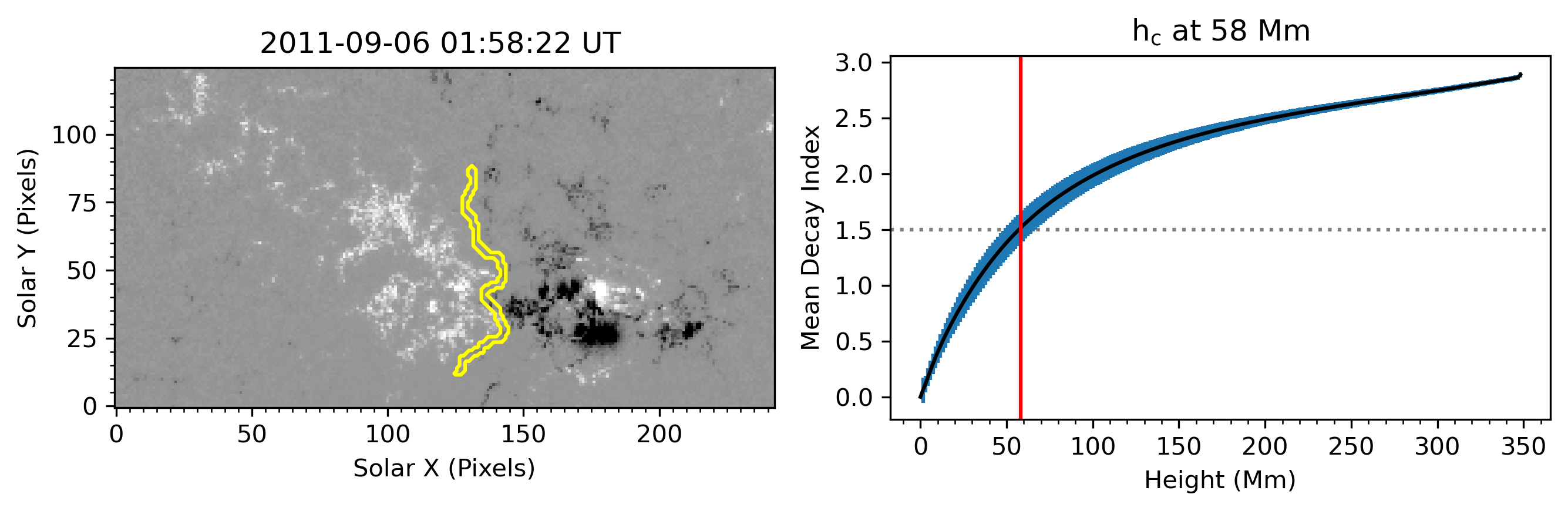}
     \end{subfigure}
     \hfill
     \begin{subfigure}
         \centering
         \includegraphics[width=0.73\textwidth]{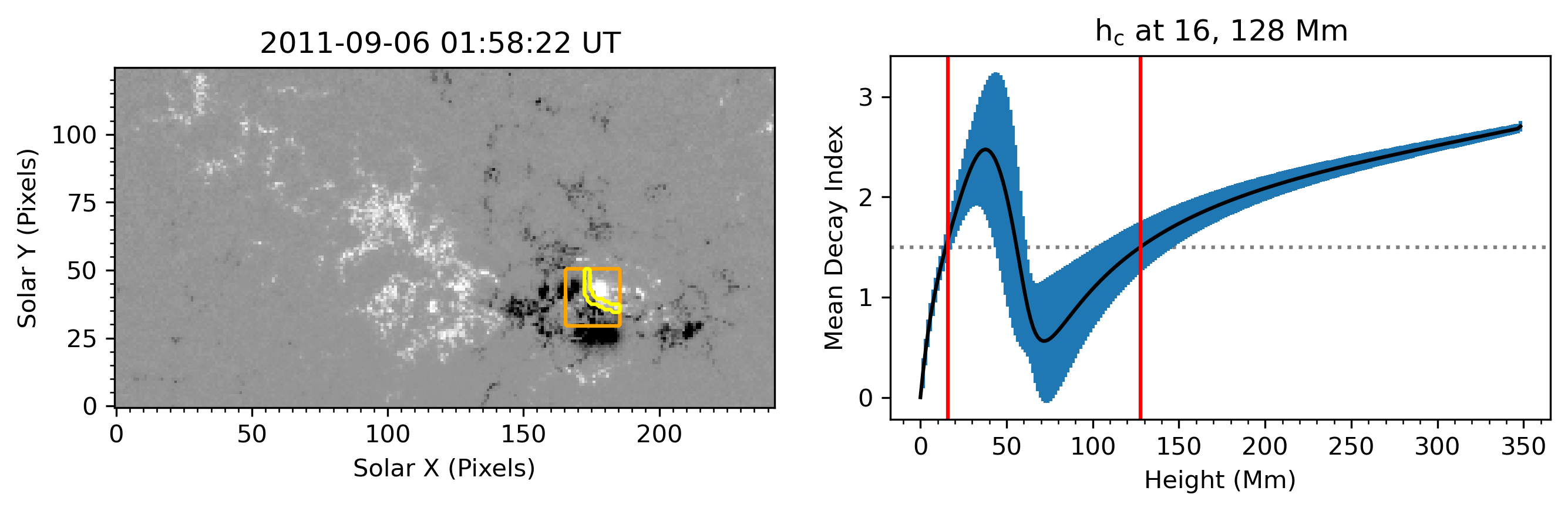}
     \end{subfigure}
        \caption{\textcolor{black}{Left: automated (top) and manual (bottom) PILs detected in NOAA AR 11283 within HARP 00833. Right: black curves represent the mean decay index vs height above the automated (top) and manual (bottom) PILs. The blue intervals represent the standard deviation of the decay index along the respective PIL at each height. The black dotted horizontal line marks the critical decay index threshold $n_{c} = 1.5$, and the red vertical lines show the corresponding critical heights. The decay index profile above the manually selected PIL exhibits a saddle-like shape with two critical heights.}}
    \label{fig:decay_index}
\end{figure*}

The differences in the methods of PIL detection used lead to differences in the critical heights we obtain.
The first method, which relies on automatically-detected PILs, is used on extrapolations of magnetograms that were taken at 00:00 UT every day of each active region’s disk passage. In this way, we can track the critical height throughout the observed disc passage of each active region (see Figure. \ref{fig:CH}). The application of this automated PIL detection to magnetograms taken at regular intervals represents a more ``operational’’ approach to estimating the critical height. However, this method has two primary limitations. 
Firstly, the automated detection of PILs can result in very long, winding PILs, along which the critical height can vary significantly. Therefore, averaging the critical heights from above every PIL pixel can lead to the inclusion of very high and very low critical heights. This is quantified by the standard deviation error estimate given with each critical height measurement. 

The second limitation is that, to estimate the critical height at the time of the CME onset, we assume the critical height (and its uncertainties) evolves linearly between the values we obtain at 00:00 UT each day and interpolate what the critical height would have been at the time of eruption. This means the extrapolation closest in time to CME onset could represent solar conditions up to 12 hours before or after the CME time. An example of this kind of interpolated critical height is shown in Fig. \ref{fig:CH} by a red `X'.

The second PIL detection method addresses these limitations by using extrapolations of magnetograms that were observed closest to the onset times of each CME. Due to the cadence of the SHARP (SMARP) data series, this means the critical heights are calculated using data no more than 12 minutes (96 minutes) before or after the CME onset time. Furthermore, our usage of EUV observations to determine the localised section of the PIL where each CME erupted from reduces the influence of potentially extreme critical height values from elsewhere along the PIL that were not physically relevant to the eruption.
Therefore, this method solves the two main drawbacks of the first, ``operational’’ method. However, unlike the ``operational’’ method, this kind of ``manual’’ critical height estimate can only be performed after a CME has already occurred. Whilst the limitations of timing could be minimised by extrapolating near-real time magnetogram datasets, predicting and selecting the specific section of PIL where a CME will erupt from is difficult. In this way, we can consider this method to give a more precise, post-event analysis of the critical height. An example of critical height obtained using this method is shown in Figure \ref{fig:CH} by a blue triangle.

\begin{figure*}[htp]
    \centering
           \includegraphics[width=\textwidth]{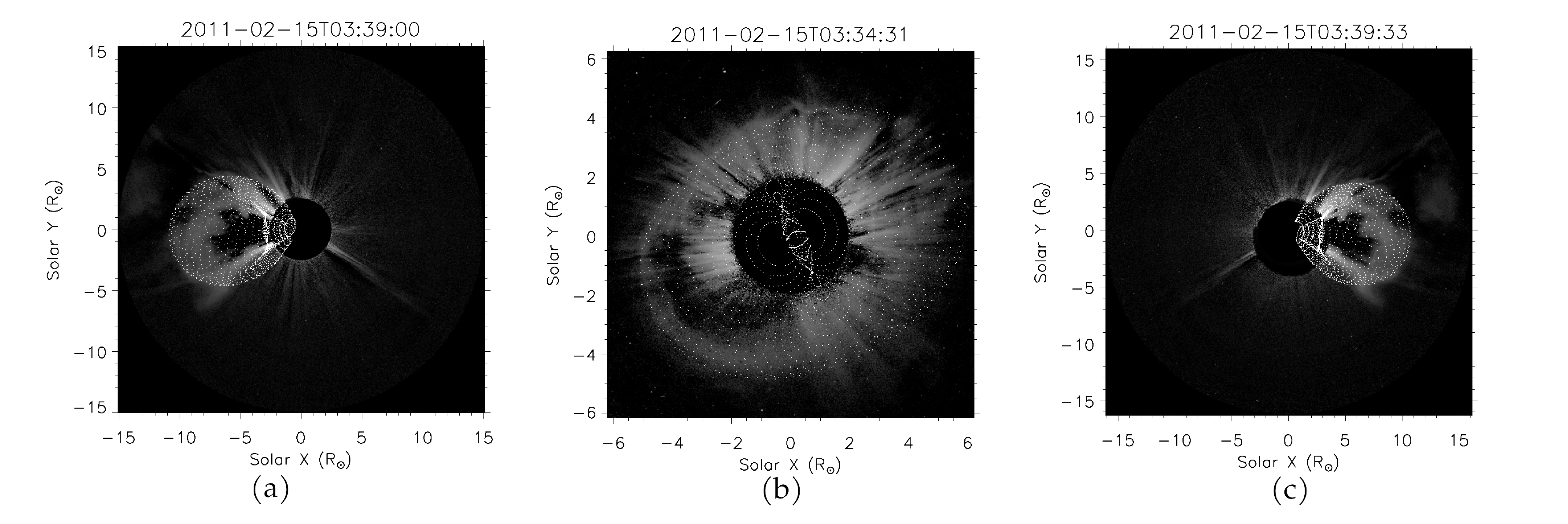}
    \caption{Fittings of the GCS flux rope to the COR2A (left), C2 (middle) and COR2B (right) images of a Halo CME of 2011/02/15 at 03:34 UTC that erupted from NOAA 11158.}
    \label{fig:GCS_fit}
\end{figure*}

Critical heights (marked with an $*$ in Table \ref{tab:table}  for selected active regions were directly extracted from \citet{james2022evolution} to increase the sample size. The method used to determine these critical heights differs slightly from the other methods employed in this study. The \citet{james2022evolution} values are based on manually defined AR subregions (not fully automated), can contain multiple PIL sections (not just the longest PIL section), and are based on magnetograms from the closest hour to each CME onset time (i.e. not interpolated). In these ways, this approach represents a hybrid between the automated and manual methods. As they are based on multiple long, messy PIL sections in contrast to the manual method which selects a single small section of a PIL where eruption signatures were observed, they have been included in the automated column of Table \ref{tab:table}.

Direct observations of EUV eruption signatures in five active regions (see table for their NOAA numbers) revealed PILs that differed from the automatically detected PILs. These manually identified PILs exhibited distinctive saddle profiles in their mean decay index variations with height. Figure. \ref{fig:decay_index} shows one such example of this type of profile. The EUV eruption signatures led us to examine a small PIL section in a different part of the active region to where the longest PIL was detected automatically. 

The decay index profiles are quite different above the two PILs. The mean decay index above the automatically detected PIL exhibits a smooth, monotonic increase with height, crossing the critical decay index threshold once at a critical height of $58\, \rm{Mm}$. The mean decay index above the PIL in the manually defined region of interest first crosses the critical decay index at a height of $16\, \rm{Mm}$ and continues to increase until about $35\, \rm{Mm}$. However, the mean decay index then begins to decrease again until a height of $70\, \rm{Mm}$, becoming sub-critical once again. Above this height, the mean decay index increases once more, reaching a second critical height at $128\, \rm{Mm}$. A torus-stable region (where the mean decay index is less than 1.5) exists above this PIL in the height range $55-128\, \rm{Mm}$.

For events with complex active region configurations, the manual method provides a more comprehensive understanding of the magnetic field's spatial variation by focusing on eruption-associated PIL regions. This approach allows the identification of features such as saddle profiles, which might otherwise be overlooked by automated methods that focus on different sections of the same active region.

\subsection{3D reconstruction and Velocity Estimation}
    \label{sec:VE}
    
3D reconstruction of 37 CMEs was performed using the GCS model. The data used for these events were sourced from the STEREO/SECCHI-COR2A and STEREO/SECCHI-COR-2B coronagraphs, STEREO Extreme UltraViolet Imager (EUVI) \citep{kaiser2008stereo}, and the SOHO/LASCO-C2/C3 \citep{bonnet1997overview} coronagraphs. EUVI and COR-2 level 0.5 data were processed to level 1.0 using the secchi\_prep.pro routine in the Solarsoft library of the Interactive Data Language (IDL). For LASCO, we utilized level 1 data, which were corrected for instrumental effects, solar north orientation, and calibrated to physical units of brightness \citep{majumdar2020connecting}.

\begin{figure}[h]
    \centering           \includegraphics[width=0.46\textwidth]{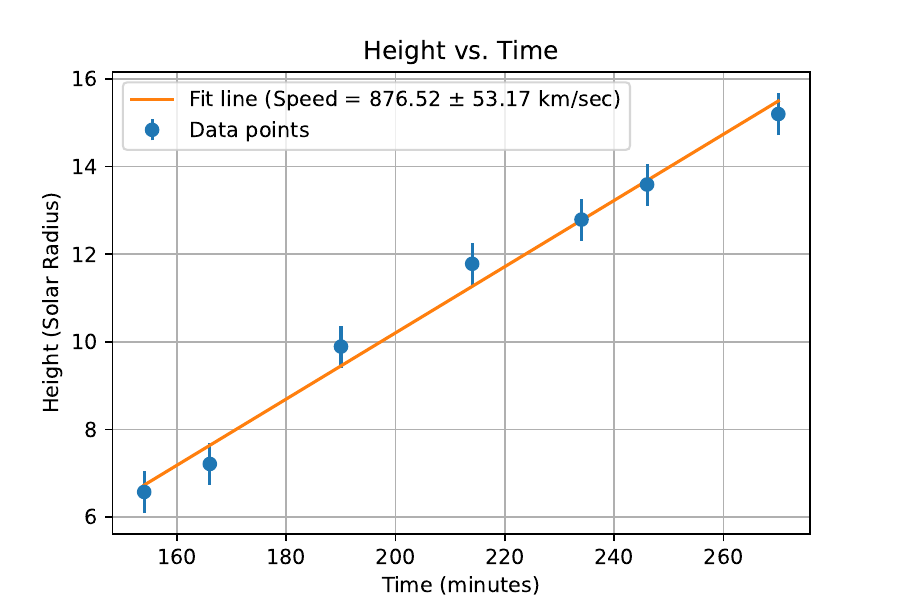}
    \caption{Height-time plot of a 2011/02/15 CME showing GCS height points in combined C2/C3, COR2A and COR2B FOV. The linear regression line represents the best fit to the data points, providing an estimate of the average 3D speed of the CME over the observed time interval.}
    \label{fig:HT}
\end{figure}

To track CME evolution in the outer corona, the GCS model was fitted simultaneously to COR-2 (field of view: 2.5–15 \Rs) and LASCO C2/C3 (field of view: 2.2–30 \Rs) images manually. An example of the GCS fitting for the event on 2011/02/15 at 03:34 UT is shown in Figure \ref{fig:GCS_fit}. Key GCS parameters such as latitude, longitude, aspect ratio, tilt angle, height, full angular width, and the derived 3D speed were recorded for each CME. Following the GCS geometrical fitting, the average 3D speed is estimated using height-time data, with a linear regression applied to determine the average 3D speed, as illustrated in Figure \ref{fig:HT}. To ensure consistency across speed estimates for all CMEs, we maintained a simple linear relationship between distance and time. As GCS fitting is a manual technique, potential human bias could influence the modeled parameters. To enhance the reliability of the speed and width estimates, we conducted the fitting procedure multiple times, exploring a range of values for each fitting parameter to reduce uncertainty.

Since three-vantage point observations have been unavailable following the malfunction of STEREO-B in 2014, certain parameters (latitude, longitude, and tilt angle) were fixed for the 2015-12-28 event based on the CME source region location, while height, half-angle, and aspect ratio were fitted to the image time series. The fitting procedure followed methodologies outlined by \citet{thernisien2006modeling, thernisien2009forward,majumdar2020connecting} and \citet{gandhi2024correcting}. 

Table \ref{tab:table} provides a comprehensive summary of the characteristics of the 37 CMEs analyzed in this study. It includes details such as CME date, first C2 appearance time, onset time, NOAA active region number, and Hale class to represent the complexity of the associated active regions. Additionally, the table lists the GCS model parameters, including derived speed and angular width, \href{https://cdaw.gsfc.nasa.gov/}{\textcolor{black}{CDAW}}-provided projected speeds, and the critical heights computed using both automated and manual PIL detection methods, along with their associated uncertainties.

Most 3D speeds were adopted from our previous work (\cite{gandhi2024correcting}), which presented a catalog \citep{harshu939_2024_cme} of 3D and 2D speeds for 360 CMEs, where these events were initially fitted, while a subset of events were newly fitted using the GCS model. 

\begin{table*}[h]
    \centering 
    \renewcommand{\arraystretch}{1.5} 
    \rotatebox{90}{ 
        \begin{minipage}{\textheight} 
        \resizebox{\textheight}{!}{ 
            \renewcommand{\arraystretch}{1.5}
        \begin{tabular}{|c|c|c|c|c|c|c|c|c|c|c|c|c|c|c|c|} \hline
        \textbf{CME No.} & \multicolumn{2}{|c|}{\textbf{First C2 Appearance Date Time [UT]}} & \textbf{CME onset time} & \textbf{NOAA} & \textbf{Hale class} & \multicolumn{5}{|c|}{\textbf{GCS reconstruction parameters}} & \textbf{CDAW speed (km/s)} & \multicolumn{2}{|c|}{\textbf{Automated critical height (Mm)}} & \multicolumn{2}{|c|}{\textbf{Manual critical height (Mm)}}  \\ \hline  \cline{2-3} \cline{7-11} \cline{13-14} \cline{15-16}
        & \textbf{Date} & \textbf{Time} & & & & \textbf{GCS Lon (deg)} & \textbf{GCS Lat (deg)} & \textbf{GCS Tilt (deg)} & \textbf{GCS Width (deg)} & \textbf{GCS Speed (km/s)} & & \textbf{Interpolated height} & \textbf{Error} & \textbf{Mean critical height} & \textbf{Error}  \\ \hline
        1 & 2009-12-16 & 04:30 & 03:12 & 11035 & $\beta$ & -2 & 7.5 & -22 & 45 & $469 \pm 46$ & 276 & 36.7 & 8.7& 32.9 & 2.0  \\ \hline
        2 & 2010-02-06 & 20:06 & 19:10 & 11045 & $\beta \gamma$ & -37.4 & -3.9 & 2 & 40 & $259 \pm 14$ & 240 & 26.8& 14.0& 30.4,343.8 & 13.0  \\ \hline
        3 & 2010-02-07 & 03:54 & 03:12 & 11045 & $\beta \gamma$ & -6 & -9 & -40.8 & 80 & $481 \pm 18$ & 421 & 29.6& 11.9& 34.8 & 0.9  \\ \hline
        4 & 2010-02-08 & 06:30 & 04:48 & 11045 & $\beta \gamma$ & 12.6 & -2.2 & -25.7 & 55 & $421 \pm 43$ & 153 & 33.8& 15.2& 49.3 & 3.0  \\ \hline
        5 & 2010-02-10 & 17:30 & 15:16 & 11046 & $\beta \gamma \delta$ & -34 & -8.9 & -6.1 & 32 & $675 \pm 51$ & 538 & 37.8& 7.1& 32.7 & 1.6  \\ \hline
        6 & 2010-02-12 & 11:54 & 09:36 & 11046 & $\beta \gamma \delta$ & -2.5 & 11 & 37 & 84 & $550 \pm 57$ & 509 & 39.1& 7.5& 29.1 & 1.3  \\ \hline
        7 & 2010-04-03 & 10:33 & 09:34 & 11059 & $\beta$ & -1 & -27 & 11 & 84 & $660 \pm 65$ & 668 & 38.4& 1.5 & 33.4 & 0.7  \\ \hline
        8 & 2010-04-08 & 04:54 & 03:12 & 11060 & $\beta$ & -3 & -6.4 & 25 & 100 & $500 \pm 75$ & 264 & 33.4 & 4.0& 30.5 & 2.1  \\ \hline
        9 & 2010-08-01 & 13:42 & 09:22 & 11092 & $\beta$ & -38 & 20 & -43 & 92 & $1260 \pm 84$ & 850 & 52.7& 20.6& 68.2 & 6.4  \\ \hline
        10 & 2011-02-13 & 18:36 & 17:34 & 11158 & $\beta \gamma$ & 4 & -6 & 12 & 44 & $592 \pm 45$ & 373 & 35.8& 13.7& 46.4 & 16.6   \\ \hline
        11 & 2011-02-14 & 08:12 & 06:46 & 11158 & $\beta \gamma$ & 7 & -5 & -65 & 46 & $480 \pm 37$ & 303 & 39.9& 15.3& 37.6 & 18.7   \\ \hline
        12 & 2011-02-14 & 14:00 & 12:46 & 11158 & $\beta \gamma$ & 12 & -8 & -37 & 43 & $494 \pm 43$ & 380 & 40.6& 15.0& 33.7 & 6.6  \\ \hline
        13 & 2011-02-14 & 18:24 & 17:22 & 11158 & $\beta \gamma$ & 15 & -6 & -45 & 38 & $627 \pm 59$ & 326 & 41.1& 14.7& 34.9 & 6.2  \\ \hline
        14 & 2011-02-15 & 02:24 & 01:58 & 11158 & $\beta \gamma$ & 16 & -10 & 36 & 140 & $876 \pm 54$ & 669 & 42.3 & 14.3 & 40.6 & 2.1  \\ \hline
        15 & 2011-03-07 & 14:48 & 14:10 & 11166 & $\beta \gamma$ & -19 & 17 & 29.6 & 35 & $965 \pm 85$ & 698 & 43.3 & 11.3 & 42.1 & 1.1  \\ \hline
        16 & 2011-08-02 & 06:36 & 04:58 & 11261 & $\beta \gamma \delta$ & 15 & 12 & 50 & 120 & $600 \pm 44$ & 712 & 32.2 & 12.6 & 27.5 & 5.9  \\ \hline
        17 & 2011-08-03 & 14:00 & 13:22 & 11261 & $\beta \gamma \delta$ & 16 & 12 & -57 & 62 & $925 \pm 73$ & 610 & 42.1 & 20.8 & 31.9 & 21.6  \\ \hline
        18 & 2011-08-04 & 04:12 & 03:58 & 11261 & $\beta \gamma \delta$ & 20 & 14 & -45 & 58.7 & $1200 \pm 95$ & 1315 & 46.1 & 22.9 & 48.1 & 34.7 \\ \hline
        19 & 2011-09-06 & 02:24 & 01:58 & 11283 & $\beta \gamma$ & 27 & 14 & -37 & 34 & $855 \pm 87$ & 782 & 47.9 & 16.4 & 19.1,127.7 & 15.9  \\ \hline
        20 & 2011-09-06 & 23:05 & 22:13 & 11283 & $\beta \gamma$ & 30 & 15 & -71 & 45 & $782 \pm 45$ & 575 & 51.9 & 19.2 & 16.9,136.4 & 10.1  \\ \hline
        21 & 2011-09-07 & 23:05 & 22:34 & 11283 & $\beta \gamma$ & 35 & 16 & 67 & 54 & $810 \pm 76$ & 792 & 48.1 & 23.4 & 15.5,124.8 & 3.6  \\ \hline
        22 & 2012-04-23 & 18:24 & 17:34 & 11461 & $\beta$ & 1.9 & -17 & 17 & 58 & $690 \pm 85$ & 528 & 39.3 & 19.8 & 25.4 & 1.2  \\ \hline
        23 & 2012-06-13 & 13:25 & 12:40 & 11504 & $\beta \gamma$ & -26 & -16 & -27 & 70 & $755 \pm 61$ & 632 & 46.2$*$ & 1.7$*$ & 48.6 & 2.3  \\ \hline
        24 & 2012-06-14 & 14:12 & 13:30 & 11504 & $\beta \gamma$ & -20 & -17 & 31 & 78 & $1020 \pm 110$ & 987 & 49.1$*$ & 1.7$*$ & 50.1 & 3.1  \\ \hline
        25 & 2012-06-30 & 07:48 & 06:07 & 11515 & $\beta \gamma$ & -40 & -21 & -43 & 45 & $423 \pm 37$ & 317 & 39.2$*$ & 18.4$*$ & 38.2 & 1.0  \\ \hline
        26&2012-07-01 &15:36 &14:57 &11515 &$\beta \gamma$ & -25 &-20 &-34 &56 & $939\pm93$ &723 &43.1$*$ &25.5$*$ &39.9 &1.6 \\ \hline
        27&2012-07-05 &13:24 &13:02 &11515 &$\beta \gamma$ & 24& -21&4 &48 &$740\pm51$ &741 &47.7$*$ &23.7$*$ &35.4 &1.6 \\ \hline
        28&2012-07-05 &22:00 &21:39 &11515 &$\beta \gamma$ &25 &-20 &-52 &56 &$990\pm83$ &980 &47.1$*$ &23.6$*$ &51.3 &3.8 \\ \hline
        29&2012-08-14 &01:25 &00:21 &11542 &$\beta \gamma$ &35 &-27 &47 &50 &$650\pm41$ &634 &44.8 &7.1 &45.1 &2.6 \\ \hline
        30&2012-11-21&16:00 &15:40 &11618 &$\beta \gamma$ &-9 &-2 &42 &102 &$680\pm75$ &529 &35.3 &22.2 &26.2 &2.8 \\ \hline
        31&2013-04-11 &07:24 &07:00 &11719 &$\beta \gamma$ &-16 &-6 &-69 &66 &$1132\pm120$ &861 &64.1 &36.0 &31.5 &1.5 \\ \hline
        32&2013-10-22 &21:48 &21:15 &11875 &  $\beta \gamma \delta$& 1& -18& -16& 55& $957\pm91$&459 &61.3 &23.0 &23.5,226.3 &1.9 \\ \hline
        33&2014-02-12 &06:00 &03:52 &11974 & $\beta \gamma \delta$ &-3.4 &9 &50 &60 &$726\pm67$ &373 &54.0 &26.3 &58.0 &25.3 \\ \hline
        34&2014-03-28 &20:00 &19:09 &12017 & $\beta$ &17 &36 &-35 &60 &$489\pm31$ &420 &46.6$*$ &9.4$*$ &20.3 &6.4 \\ \hline
        35&2014-03-28 &23:48 &23:44 &12017 &$\beta$ &16 &15 &-42 &64 &$640\pm71$ &514 &47.4$*$ &10.4$*$ &27.8 &12.5 \\ \hline
        36&2014-03-30 &12:24 &11:45 &12017 &$\beta$ &15 &15 &-36 &92 &$620\pm63$ &487 &48.4$*$ &18.4$*$ &32.3 &18.3 \\ \hline
        37&2015-12-28 &12:12 &11:20 &12473 &$\beta \gamma$ &6 &-13 &-38 &102 &$1395\pm135$ &1212 &50 .3&10.3 &57.6 &0.8 \\ \hline
        \end{tabular}
                }
        \vspace{0.1cm}
        \caption{
       List of CMEs with GCS reconstruction parameters and critical height measurements. The 'First C2 Appearance Date Time [UT]' includes the initial appearance date and time in C2. 'GCS reconstruction parameters' cover source location, tilt, width, and 3D speed derived from the GCS model. 'Automated critical height' and 'Manual critical height' present the interpolated and mean critical heights with associated errors. Critical heights (and its uncertainties) with an $*$ are taken from \citet{james2022evolution}   
        }
        \label{tab:table}   
        \end{minipage}
        }
\end{table*}

\section{Results}
    \label{sec:Results}
This section presents the results obtained using automatic (Method I) and manual (Method II) PIL detection approaches and their relationship to CME speeds. Critical heights, which represent the altitudes where torus instability is triggered, are calculated using radialized-field magnetograms from the SMARP and SHARP databases. Projected (2D) speeds are sourced from the SOHO LASCO CDAW catalog, and 3D speeds obtained through the GCS model applied to multi-point coronagraph data. Confidence intervals (CI) for linear/Pearson (R) and rank-order/Spearman ($\rho$) correlation coefficients are computed using the bootstrapping method to ensure robust statistical estimates. By evaluating the relationship between critical heights and CME speeds across these methodologies, the analysis highlights the influence of PIL detection techniques on understanding CME dynamics and provides insights into the accuracy of magnetic field-based predictions for space weather forecasting.

\subsection{2D and 3D CME speeds vs critical heights from automatic PIL detection}

\begin{figure}[h]
    \centering
    \includegraphics[width=0.46\textwidth]{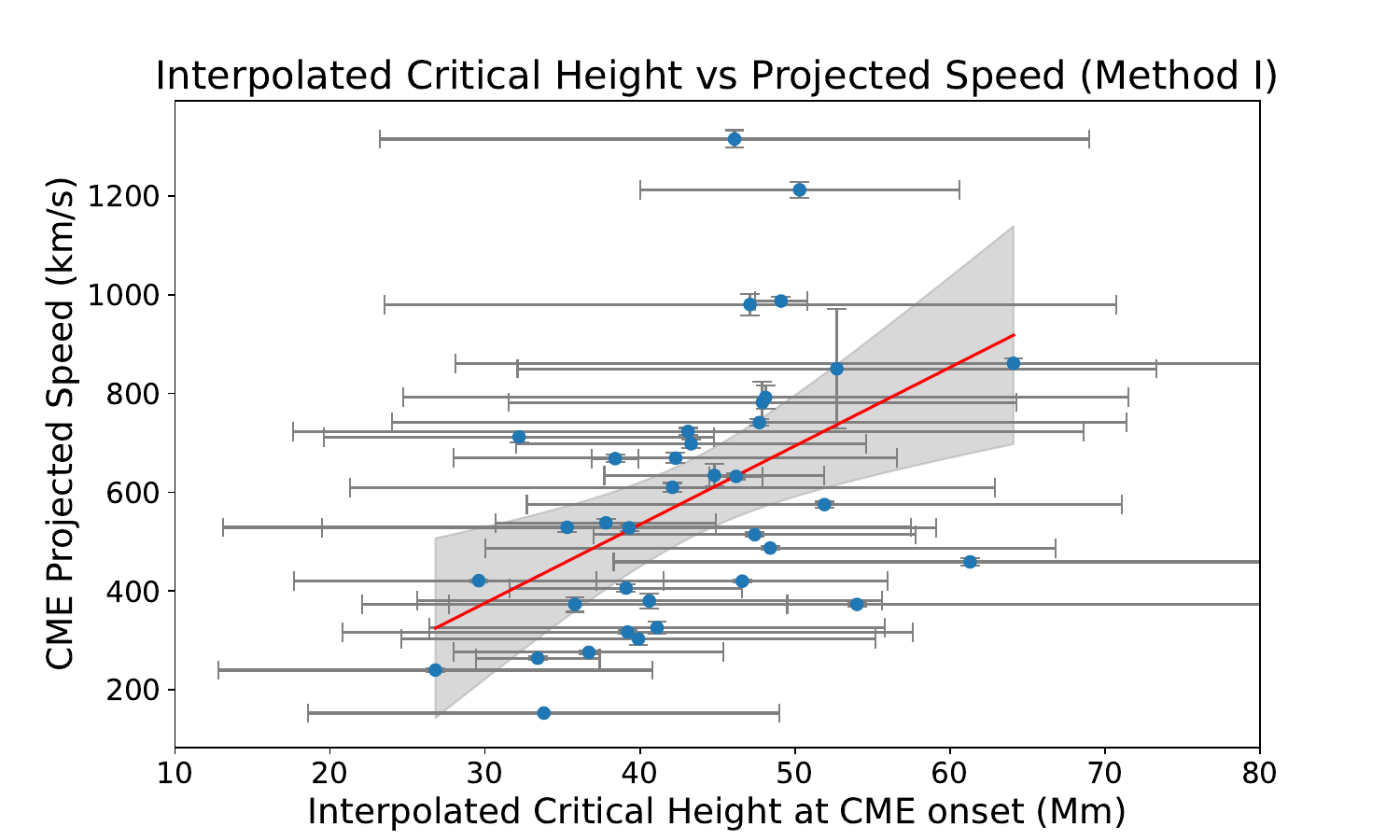}
    \caption{Interpolated critical height at CME onset (Mm) plotted against CME projected speed (km/s) for 37 events. The solid line represents the linear fit ($y = 15.34x - 71.79$) with a 95\% confidence interval. The linear/Pearson correlation coefficient is $R = 0.48$ $\pm$0.12 and Spearman correlation coefficient is $\rho =0.55$ $\pm$ 0.15.}
    \label{fig:fig6}
\end{figure}

Figure \ref{fig:fig6} shows the relationship between interpolated critical heights and 2D CME speeds. Interpolated critical heights are derived from automatic polarity inversion line (PIL) detection, averaged along the PIL, and calculated by assuming linear evolution between daily magnetogram observations at 00:00 UT. A total of 37 CMEs are analyzed. The linear/Pearson correlation between critical height and 2D speed is  $R = 0.48$ $\pm$ 0.12 (1$\sigma$), with a 68\% CI of (0.36,0.61) and a statistically significant $p<0.001$. The Spearman correlation coefficient, which is more robust to outliers is $\rho =0.55$ $\pm$ 0.15 (1$\sigma$), with a 68\% CI of (0.40, 0.69) and $p<0.001$. The later is numerically higher than the former. However, their overlapping confidence intervals suggest that the difference is not statistically significant. The fitted linear regression model,  $y = 15.34x - 71.79$ , shows a slope of 15.34 km/s per Mm. However, variability is evident, as shown by the confidence intervals around the fitted line, attributed to the influence of averaged critical heights from the long, winding PILs.

\begin{figure}[h]
    \centering
    \includegraphics[width=0.46\textwidth]{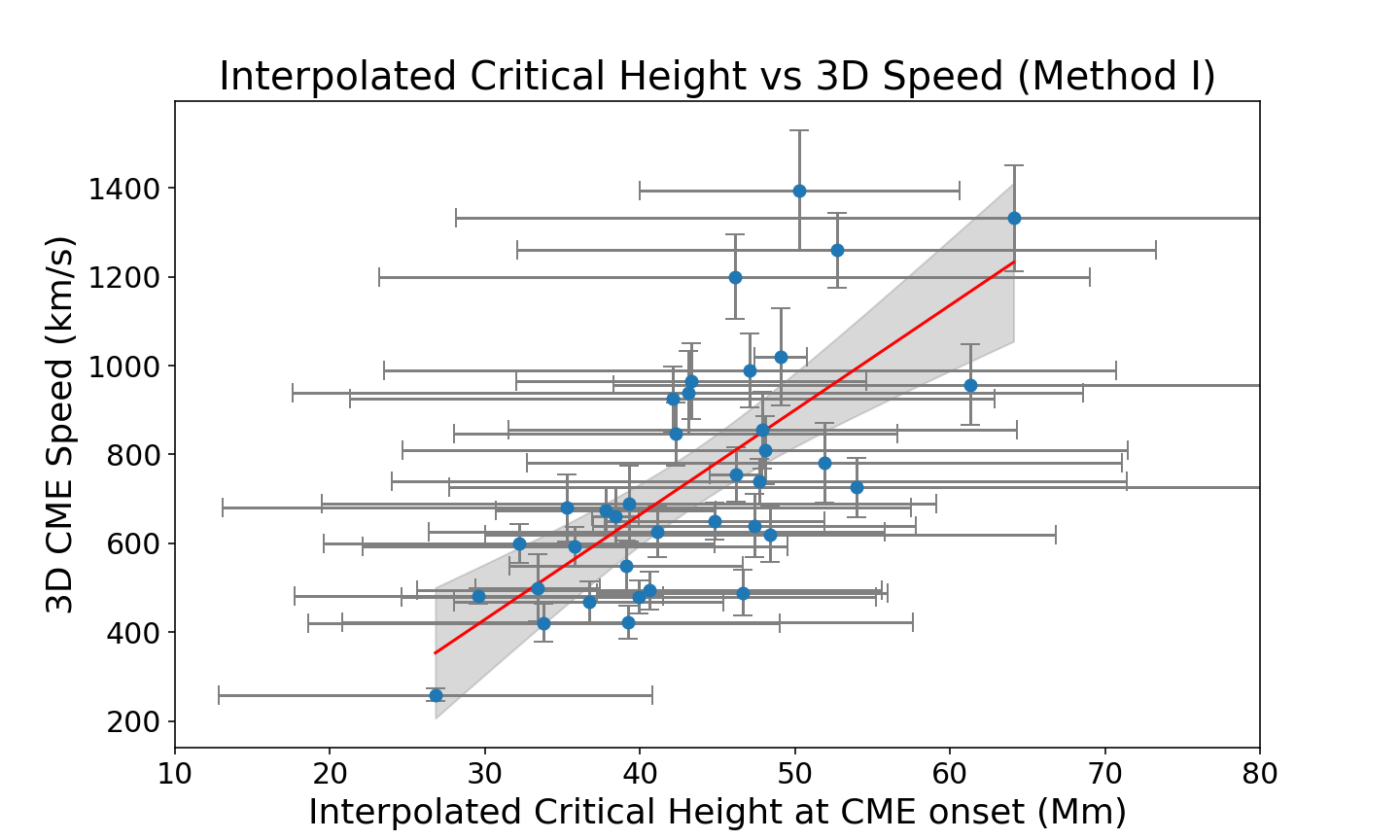}
    \caption{Interpolated critical height at CME onset (Mm) plotted against 3D CME speed (km/s) for 37 events. The solid line represents the linear fit ($y = 23.57x - 278.52$) with a 95\% confidence interval shown as grey shaded region. The Pearson and Spearman correlation coefficient is 0.71 $\pm$ 0.08.}
    \label{fig:fig7}
\end{figure}

Figure \ref{fig:fig7} presents the relationship between interpolated critical heights and 3D CME speeds derived from the GCS method using multi-point coronagraph data. The Pearson correlation between interpolated critical heights and 3D speeds is $R = 0.71$ $\pm$ 0.08, with a 68\% CI of (0.62, 0.78) and a statistically significant ( $p < 0.001$ ). The Spearman correlation coefficient is $\rho = 0.71$ $\pm$ 0.08 (1$\sigma$), with a 68\% CI of (0.62, 0.78) and $p<0.001$. Both correlations are identical, indicating strong agreement between the two measures. The linear regression model,  $y = 23.57x - 278.52$, shows a steeper slope of 23.57 km/s per Mm. This result reflects a more substantial increase in CME speed per unit rise in critical height when using 3D speeds, as compared to the 2D speeds. The tighter clustering of data points around the regression line suggests that 3D CME speed aligns more closely with the critical height compared to the projected speed analysis.

These results highlight the stronger correlation between interpolated critical heights and 3D CME speeds compared to projected speeds. While the increased number of pixels can lead to larger standard deviations and the difference in time between the extrapolations and the CME onset introduces inaccuracy, the method still captures statistically significant trends in the relationship between critical heights and CME dynamics, reinforcing the method’s reliability in representing key magnetic conditions that influence CME kinematics. 

\subsection{3D CME speeds vs Mean critical heights from manual PIL detection}

\begin{figure}[H]
    \centering
           \includegraphics[width=0.46\textwidth]{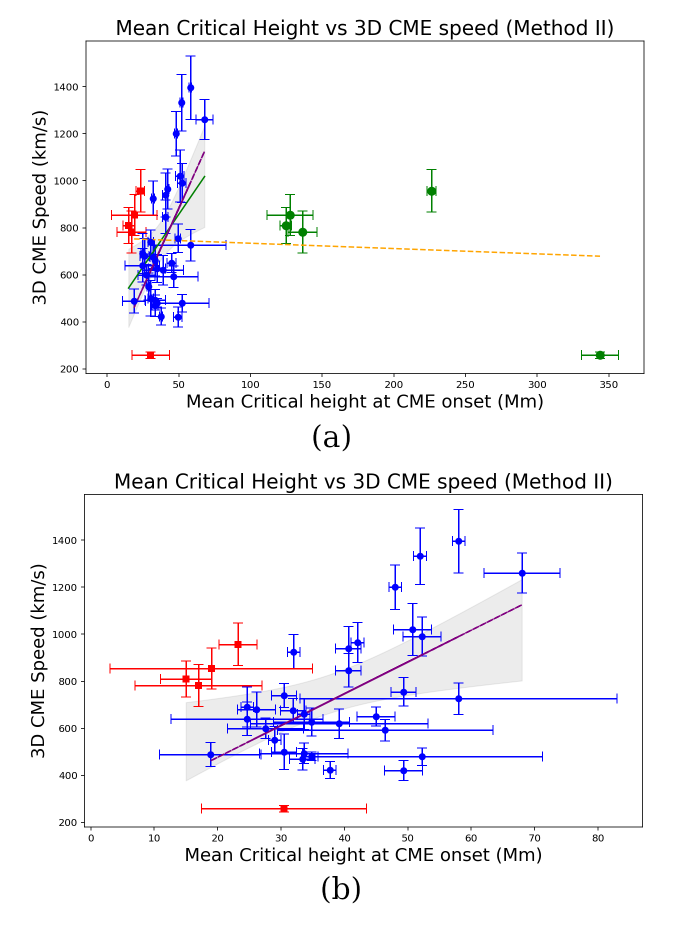}
    \caption{Mean critical height at CME onset (Mm) vs. 3D CME speed (km/s) with error bars is shown for 37 events in (a), with blue points for 32 non-saddle CMEs, and red/green for saddle events with lower/higher critical heights. (b) excludes green points. The violet line fits blue points only ($R = 0.58$$\pm$0.13 and $\rho = 0.42$ $\pm$0.18), the green line includes blue and red points ($R = 0.43$$\pm$0.15 and $\rho = 0.27$ $\pm$0.18),and the yellow dashed line indicates almost zero correlation.}
       \label{fig:8}
\end{figure}

The relationship between mean critical heights and 3D CME speeds, as determined using manually selected polarity inversion lines (PILs), is shown in Figure \ref{fig:8}. Unlike the automated method, the manual method focuses on identifying PILs within sub-regions of the active region along which the CME originated. This approach utilizes EUV observations to refine the selection of the PIL section, allowing for more localized critical height measurements. 

Figure \ref{fig:8} presents data for all 37 CMEs analyzed, with blue points representing the 32 non-saddle events (single critical height) and red and green points indicating the five saddle events (two critical heights). The correlation between mean critical height and 3D CME speed varies based on how events with saddle-shaped decay index profiles (hereafter referred to as saddle events or saddle profiles) are treated. When saddle events are excluded, the Pearson correlation coefficient is $R=0.58$$\pm$0.13, with a 68\% CI of (0.43, 0.69) and the Spearman correlation is $\rho=0.42$$\pm$0.18, with a 68\% CI of (0.23, 0.58), indicating a moderate linear relationship between mean critical height and CME speed compared to the relationship from Method I. The linear fit shown in violet is described by y = 13.50x + 206.97 Including saddle events by considering the lower critical heights (red points) reduces the correlations to $R=0.43$$\pm$0.15, with a 68\% CI of (0.26, 0.56) and $\rho=0.27$$\pm$0.18, with a 68\% CI of (0.08, 0.44) with the linear fit given by y = 8.94x + 409.87, reflecting that the inclusion or exclusion of saddle profiles affects the overall spread in the mean critical heights. Including saddle events by considering the higher critical height (green points) results in a near-zero correlation (yellow dashed line), suggesting a poor relationship between these heights and CME speeds.

The scatterplot demonstrates the impact of saddle events on the overall correlation. The red points, corresponding to lower critical heights, align partially with the trend of non-saddle events but introduce substantial variability, weakening the correlation. The green points, representing higher critical heights, deviate significantly from the main trend, resulting in negligible or negative correlation values. These results highlight the sensitivity of the manual method to the choice of critical height in regions with complex decay index profiles. Among the five saddle CMEs, the differences between lower and higher critical heights have a notable impact on the overall trends observed in the data.

\section{Discussion and Conclusion}
    \label{sec:D&C}

This study explores whether there is a relationship between mean critical heights and CME speeds, focusing on automated and manual polarity inversion line (PIL) detection methods. The automated method, which analyzes the longest PIL structure in case of multiple PILs in an active region, demonstrates a strong positive linear correlation ($R = 0.71$ $\pm$ 0.08) with 3D CME speeds as shown in Figure \ref{fig:fig7}. This result highlights its reliability for capturing large-scale magnetic conditions influencing CME dynamics. The manual method, while providing localized insights by focusing on subset of pixels relevant to eruption events, achieves a lower linear correlation ($R=0.58$$\pm$0.13) for non-saddle events as shown in Figure \ref{fig:8}.

Saddle profiles (see bottom panel of Figure \ref{fig:decay_index}), characterized by multiple critical heights above the same PIL section, are a significant source of complexity, as observed in the manual method. When lower critical heights (red points) are included, the linear correlation decreases to  $R=0.43$$\pm$0.15 . Including higher critical heights (green points) results in a negligible correlation, highlighting their deviation from the main dataset dominated by lower critical heights as shown in Figure \ref{fig:8}. These findings strongly support the choice of using the lower critical height in saddle regions as a more reliable predictor of CME speed. This result validates the approach of prioritizing lower critical heights and represents a significant contribution to the field, providing a robust basis for future studies. Combining this result with automated PIL detection offers a promising framework for improving pre-eruptive predictions.

Interestingly, no saddle profiles were observed in the automated PIL detection results. This could be due to the averaging across longer, continuous PILs, which smooths out localized variations in the decay index profile. Longer PILs, while reducing the likelihood of identifying saddles, introduce other forms of variability due to fluctuations along their length. This observation aligns with previous studies that emphasize the importance of long PILs in capturing overall magnetic stability (e.g., \citet{xu2012relationship}). However, further investigation is needed to determine the occurrence and influence of saddles in both automated and manual methods.

The findings presented in this work are consistent with our hypothesis that higher critical heights correlate with higher CME speeds because the weaker coronal magnetic fields above the onset of the torus instability (nc(h) = 1.5) allow the flux ropes to accelerate more effectively after reaching the torus instability threshold. The need for the flux rope to rise further before reaching the critical height might allow increased energy build-up, contributing to higher CME speeds. The strong linear/Pearson and Spearman positive correlation observed using the automated method $R = 0.71$ $\pm$ 0.08 supports this interpretation, as longer PILs analyzed by this method provide a more comprehensive view of the magnetic environment above a source region, influencing CME dynamics. The lower correlation in the manual method $R=0.58$ $\pm$ 0.13 for non-saddle events highlights the localized variations in magnetic conditions that may affect the flux rope’s acceleration, emphasizing the complexity of the critical height-speed relationship. This study thus provides empirical evidence supporting the role of critical height as a diagnostic parameter for predicting CME speeds.

An alternative hypothesis is that faster CMEs can result from regions with lower critical heights. This view suggests that eruptions in these regions could involve more flare reconnection as the flux rope rises through stronger overlying magnetic fields, hence, accelerating the CME. However, our results do not support this hypothesis. Instead, the strong correlations we observe between critical height and CME speed suggest that both the weaker overlying coronal fields above the torus instability zone and increased buildup of energy in the flux rope at higher critical heights are the factors affecting CME speeds. Alternatively, the stronger underlying fields could raise the critical height by forcing the CME outward more forcefully. Future observational and modeling studies should aim to distinguish between these two scenarios by measuring both the overlying field strength and the underlying field strength near the eruption site, which would help clarify the roles of these competing (or coexisting) factors.

The findings presented here build on prior work, such as \citet{james2022evolution}, which identified the critical decay index as a determinant for CME initiation. While previous studies have often relied on projected speeds in relation to decay index, this study demonstrates that 3D speeds derived from GCS model correlate more strongly with critical heights ( $R = 0.71$ $\pm$ 0.08 vs. $R = 0.48$ $\pm$ 0.12 shown in Figures \ref{fig:fig6} and \ref{fig:fig7}). This improvement emphasizes the necessity of multi-point observations for reducing projection effects, a limitation inherent in single-viewpoint data. The improved correlation highlights the potential of critical height as a predictive metric when combined with 3D speed measurements, advancing our understanding of CME dynamics.
 
The automated PIL detection method emerges as a valuable tool for operational forecasting, offering high correlations with CME speeds while maintaining efficiency for large-scale monitoring. Although daily interpolations introduce timing limitations, the integration of near-real-time magnetograms could mitigate this issue, enabling forecasters to update critical height measurements closer to CME onset. This approach bridges the gap between precision and practicality, addressing the primary limitation of interpolation-based methods.

The manual method, while less suited for real-time applications in terms of time consumption, offers detailed insights into localized PIL dynamics and the impact of saddle profiles. Future improvements in combining these approaches could leverage the strengths of both methods, enabling accurate, timely predictions. For example, integrating automated PIL detection with manual refinements in high-risk regions could enhance forecasting accuracy without sacrificing operational efficiency.

The results presented here lay the foundation for exploring additional factors influencing CME dynamics. While this study focuses on critical heights as a predictor of CME speeds; magnetic field strength ($|B|$) and other reconnection-related parameters have been identified as important factors in previous studies \citep{moon2002statistical,qiu2005magnetic,guo2006quantitative,guo2007magnetic,jain2010relationship,bein2012impulsive,berkebile2012cme,salas2014statistical,takahashi2016scaling}. Testing predictions that CME speed scales with the product of ($|B|$) and reconnected flux, as well as examining their correlations with decay index, represents a natural extension of this work. Previous investigations into the formation of fast CMEs have highlighted various source-region parameters that correlate with CME speeds. \citet{su2007determines} and \citet{wang2008statistical} found weak but positive correlations between CME speeds and extensive parameters such as magnetic flux, area, and average photospheric field strength. Structural properties, including the number of polarity inversion lines (PILs), their length, and measures of magnetic complexity, also show varying levels of correlation \citep{guo2006quantitative, kontogiannis2019photospheric}. Notably, \citet{kontogiannis2019photospheric} reported that the length of the main PIL and total non-neutralized current exhibited high correlation coefficients ($c>0.8$) for fast CMEs ($Vcme > 750km/s$).

Combined parameters, which account for both the magnitude and structure of the source region, have been identified as stronger predictors of CME speeds. Free magnetic energy and helicity, as indicators of non-potentiality, show moderate-to-strong correlations with CME speed \citep{venkatakrishnan2003relationship, liu2007speed, gopalswamy2009cme}. For example, \citet{park2012occurrence} and \citet{kim2017relation} found high correlations (
$c=0.8$) between helicity injection rates and CME speeds in cases with a uniform helicity sign during the energy buildup phase. \citet{xu2012relationship} and \citet{deng2017roles} demonstrated that a higher decay index, indicative of reduced downward tension forces in the coronal magnetic field, correlates well with higher CME speeds, with \citet{xu2012relationship} suggesting that this parameter underpins several other source-region effects.

The relationship between CME speed and flare reconnection parameters has also been explored extensively. \citet{deng2017roles} observed a correlation coefficient of 0.76 for ribbon flux, which quantifies reconnected magnetic flux during the associated solar flare, in a sample dominated by fast CMEs. However, other flare-related measures, such as peak soft X-ray flux and reconnection rate, typically exhibit weaker correlations, emphasizing the need to consider additional physical factors. \citet{liu2007speed} further highlighted that CMEs originating from unidirectional open-field structures (pseudostreamers) tend to be faster on average than those from closed-field loop arcades.

Incorporating these findings into models, along with the relationships between decay index,
$|B|$, and reconnected flux, could significantly improve predictive capabilities. However, the scatter in many of these correlations indicates that CME speed is likely influenced by a complex interplay of multiple parameters, including both primary source-region properties and secondary effects such as the coronal field structure and solar wind interactions

This study demonstrates the predictive potential of mean critical height for CME speeds but acknowledges several limitations. The reliance on the GCS geometrical model assumes a three-part CME structure, which may not apply to all events. Similarly, while the automated method efficiently captures large-scale variations, it overlooks small-scale variations that could affect specific CME events.

 In five cases (one such case shown in Figure \ref{fig:decay_index}), the longest-contiguous PIL as identified by Method I is not associated with the EUV emission corresponding to the eruption of interest. This may suggest that such PILs, while dominant in the photospheric field, do not meaningfully contribute to the dynamics of the eruption. Nevertheless, critical height as determined using the longest-contiguous PILs provides very strong correlations with CME speed, indicating the robustness of this method. This would suggest that dynamically significant critical heights may be insensitive to the detailed photospheric structures, such as the exact choice of PILs, but rather it may be enough to capture the overall decay of the magnetic field of the source region with height in order to predict the CME speeds. This is in agreement with expectations that higher-order multipole terms, corresponding to finer-scale photospheric magnetic fields, fall off more quickly with height to leave the largest-scale photospheric fields dominating the extrapolated magnetic structure at active-region-scale altitudes \citep{archontis2019emergence,liu2020magnetic,vemareddy2024magnetic}. Again, this underlines the importance of large-scale magnetic configurations over the fine details of individual PILs in determining critical heights. Further investigation of this insensitivity could improve the calculation of critical heights for operational forecasting.

The manual method, while providing detailed insights into localized PIL spatial variations, is less suited for operational use due to its time-intensive nature. The variability introduced by saddle profiles in critical height estimation highlights a complexity that may not always be apparent in the automated method, potentially due to its averaging process or the limitations of the dataset analyzed here. Further investigation with a larger dataset is needed to confirm whether saddle profiles are consistently absent in automated detections or if they are simply smoothed out along the longer PILs.

Future research should address these limitations by expanding datasets to include a wider range of CME events and refining critical height detection methods before operational forecasting can be done. While the critical height is essential for understanding CME dynamics, it is also important to consider that other factors, such as solar cycle variations and magnetic field strengths, can significantly influence CME behavior and characteristics. These efforts would advance space weather forecasting by providing a more comprehensive framework for understanding CME dynamics. 

Our results demonstrate that for 37 CMEs, the near real-time automated calculation of the critical height for active regions can provide a rough estimate of the speed of any CME that may erupt, which is evidently beneficial for forecasting. This initial speed estimate can also complement or serve as a useful starting point for subsequent speed measurements obtained from coronagraph observations.

\section*{Acknowledgments}
H.G. is supported by the Science and Technology Facilities Council (STFC) PhD award at Aberystwyth University. A.W.J. acknowledges funding from the STFC Consolidated Grant ST/W001004/1. The authors gratefully acknowledge the SECCHI/STEREO consortium for providing the data used in this study. We also thank the SOHO/MDI, SOHO/LASCO, and SDO/HMI and EUV consortia for their invaluable data contributions. The authors extend their gratitude to L.M.G and H.M. for their constructive comments, which greatly enhanced this study. Without their input, this work would not have been possible. We also acknowledge FBAPS, Aberystwyth University, for providing computing facilities and support. Finally, we express our gratitude to all contributing teams for their dedication to supporting open scientific research.
\newpage
\balance
\bibliography{references}
\bibliographystyle{aasjournal}



\end{document}